\documentclass[sigconf]{acmart}

\usepackage{graphicx}
\usepackage{amsmath}
\usepackage{multirow}
\usepackage[flushleft]{threeparttable}
\usepackage{subfig}
\usepackage{placeins}
\usepackage{stfloats}
\usepackage{color}
\usepackage{svg}
\usepackage{cleveref}
\usepackage{tablefootnote}
\usepackage{balance}

\usepackage{booktabs} 

\copyrightyear{2020}
\acmYear{2020}
\setcopyright{acmlicensed}
\acmConference[CHIIR '20]{2020 Conference on Human Information Interaction and Retrieval}{March 14--18, 2020}{Vancouver, BC, Canada}
\acmBooktitle{2020 Conference on Human Information Interaction and Retrieval (CHIIR '20), March 14--18, 2020, Vancouver, BC, Canada}
\acmPrice{15.00}
\acmDOI{10.1145/3343413.3377960}
\acmISBN{978-1-4503-6892-6/20/03}

\begin{document}
\fancyhead{} 

\title[Relevance Prediction from Eye-movements using CNNs]{Relevance Prediction from Eye-movements Using Semi-interpretable Convolutional Neural Networks}

\author{Nilavra Bhattacharya}
\orcid{0000-0001-7864-7726}
\affiliation{%
  \department{School of Information}
  \institution{The University of Texas at Austin}}
\email{nilavra@ieee.org}

\author{Somnath Rakshit}
\orcid{0000-0002-2017-5997}
\affiliation{%
  \department{School of Information}
  \institution{The University of Texas at Austin}}
\email{somnath@utexas.edu}

\author{Jacek Gwizdka}
\orcid{0000-0003-2273-3996}
\affiliation{%
  \department{School of Information}
  \institution{The University of Texas at Austin}}
\email{chiir2020@gwizdka.com}

\author{Paul Kogut}
\affiliation{%
  \department{Rotary and Mission Systems}
  \institution{Lockheed Martin Corporation}}
\email{paul.a.kogut@lmco.com}

\renewcommand{\shortauthors}{Bhattacharya, Rakshit, Gwizdka, and Kogut}

\begin{abstract}

We propose an image-classification method to predict the perceived-relevance of text documents from eye-movements.
An eye-tracking study was conducted where participants read short news articles, and rated them as relevant or irrelevant for answering a trigger question.
We encode participants' eye-movement scanpaths as images, and then  train a convolutional neural network classifier using these scanpath images.
The trained classifier is used to predict participants' perceived-relevance of news articles from the corresponding scanpath images.
This method is content-independent, as the classifier does not require knowledge of the screen-content, or the user's information-task.
Even with little data, the image classifier can predict perceived-relevance with up to 80\% accuracy.
When compared to similar eye-tracking studies from the literature, this scanpath image classification method outperforms previously reported metrics by appreciable margins.
We also attempt to interpret how the image classifier differentiates between scanpaths on relevant and irrelevant documents.

\end{abstract}

%
%




%


\maketitle

\section{Introduction}
\label{sec:intro}

\textit{Information relevance} is one of the fundamental concepts in Information Science in general, and Information Retrieval (IR) in particular \cite{saracevic2007relevance_review, saracevic2016notion}.
The primary purpose of IR systems is to fetch content which is useful and relevant to people.
Understanding the cognitive processes of even one individual is challenging enough, and IR systems have to cater to a variety of users, who may have wildly different mental models of what they consider to be useful and relevant.
To add another layer of complexity, these mental models are not static.
They evolve as users' knowledge and information needs change.
Researchers have investigated various forms of `signals' generated by users interacting with IR systems, that can serve as proxies for their mental processes.
Examples include search queries, mouse-clicks, logs of viewed documents, and other forms of interaction-data.
These proxies have been studied to infer what kind of information is relevant to users' needs.
Efforts from a system-centred perspective have been towards minimizing the gap between the users' query and the documents retrieved. 
The search query is considered to be an exact representation of the users' information needs. 
Documents matching the query using a given algorithm are deemed to contain the information that users are searching for, and are therefore relevant.
This notion of relevance is regarded as \textbf{algorithmic-}, or \textbf{system-relevance} \cite{saracevic2016relevance}.
The limitation of this perspective is that the query is seldom an exact representation of what the user is looking for.
As a result, retrieved documents often do not satisfy the user's information needs.

In a human-centred perspective, relevance arises from interactions between a user's information need and information objects \cite{borlund2003concept}.
This interaction results in several manifestations of relevance \cite{saracevic2016relevance}, and becomes meaningful ``only ... in relation to goals and tasks'' \cite{hjorland2010foundation}.
Our interest is in \textit{situational relevance}, or \textit{utility}.
As introduced by Wilson, ``situationally relevant items of information are those that answer, or logically help to answer, questions of concern'' \cite{wilson1973situational}.
In this paper, we refer to situational relevance as the users' \textbf{perceived-relevance} of the documents they examine for answering a question.


Neuro-physiological methods provide an interesting avenue to observe users while they interact with information systems.
One popular method is eye-tracking. 
It captures the eye-movement patterns of users as they examine information on a screen.
Eye-tracking has been frequently used to assess if the screen-content is relevant to the user (Section~\ref{sec:bg_relevance_et}).
The method has some distinct advantages.
Eye-tracking is non-invasive, and requires minimum to no effort from the user. 
Even when users are not clicking the mouse or typing a query, they are viewing the screen, and thus helping to provide continuous data in a more natural setting.
Eye-tracking can give insights about the focus and progression on an information searcher's attention in real-time.
Eye-movements are sometimes considered to be a closer proxy for human cognition \cite{just_psychology_1987}, than queries and interaction logs.


Despite its many advantages, interpreting eye-tracking data is not straightforward.
Often, a variable-length stream of real numbers are collected per stimulus. 
For the dearth of standard methods, researchers resort to aggregating this data-stream into a set of single numbers, or \textit{features}, at various levels of analysis (stimulus level, trial level, and/or participant level).
By collapsing the eye-tracking data in this fashion, the fine grained information about the individual user's progress is lost. 
This reduces the robustness and generalizability of insights gained from the analysis.


We propose an image-classification method to predict user's perceived-relevance from their eye-movement patterns.
Our method is free from many of the inherent problems associated with analyzing eye-tracking data, as shown in existing literature (Section \ref{sec:bg}).
Specifically, we convert participant's eye-movement scanpaths into images (Section \ref{sec:encoding_scanpaths}), and then transform the relevance-prediction problem into an image-classification problem (Section \ref{sec:analysis_cnn}).
For this purpose, we use state-of-the art image classifiers based on convolutional neural networks.
Our method gives promising results, and outperforms many previously reported performances in similar studies by appreciable margins (Section \ref{sec:res_classifier}).
We also attempt to interpret how the classifier possibly differentiates between user-reading-patterns on relevant and irrelevant documents (Section \ref{sec:res_interpretability}).
Finally, we discuss the limitations of our approach, and propose future directions of research (Section \ref{sec:conclusion}).




\vspace{-.75em}
\section{Related Work}
\label{sec:bg}

\subsection{Information Relevance and Eye-tracking}
\label{sec:bg_relevance_et}

One of the earliest studies employing eye-tracking for inferring users' perceived-relevance was reported by Saloj{\"a}rvi et al. \cite{salojarvi2005ImplicitRelevanceFeedback}.
Participants saw a question and a list of ten sentences.
One sentence had the correct answer to the question, and the others were either relevant or irrelevant to the question. 
Hidden Markov Models were used to predict the type of sentences the participants were reading.
Many subsequent studies have investigated the relationship between eye-movements and viewing relevant vs. irrelevant information.
These studies employed similar experimental setups, where participants examined a list of words, sentences, or documents, and judged their relevance in relation to a specific query or task.

In a majority of these relevance assessment studies, a common theme is to collapse the stream of eye-movement data into a set of single-number features, at various levels of analysis (stimulus, trial, or participant level). 
These features are then used for statistical inferences, classification, and prediction.
For instance, some variants of aggregated fixation-count and fixation-duration were used in studies reported in \cite{puolamaki2008LearningLearnImplicit, fahey2011DocumentClassificationRelevance, loboda2011InferringWordRelevance, frey2013DecisionmakingInformationSeeking, gwizdka2014CharacterizingRelevanceEyetracking, gwizdka2017DifferencesReadingWorda, wittek2016RiskAmbiguityInformation, wenzel2017RealtimeInferenceWord}.
Eye-dwell time and/or visit time was used by Fahey et al. \cite{fahey2011DocumentClassificationRelevance}.
Saloj{\"a}rvi and colleagues identified a comprehensive list of 22 such features \cite{salojarvi2005inferring}, which were later used by others (e.g., Hardoon et al. \cite{hardoon2007InformationRetrievalInferring}).

While fixation-count, fixation-duration, and dwell-time are generic eye-movement features applicable to any type of stimuli, several studies used specific features for reading text.
These works first labelled each eye-fixations as either reading or scanning/skimming.
Then they used derived measures from these two types of fixations.
Buscher et al. \cite{buscher2008EyeMovementsImplicit} used reading-to-skimming ratio to infer when participants were reading relevant text.
Over a group of studies, Gwizdka et al. \cite{gwizdka2014CharacterizingRelevanceEyetracking, gwizdka2014NewsStoriesRelevance, gwizdka2017DifferencesReadingWorda, gwizdka2017TemporalDynamicsEyetracking} reported that 
reading speed, 
number of fixations on words, 
count and length of reading sequences, 
count and percentage of words fixated upon,
durations of reading and scanning, 
and
distance covered by scanning 
proved to be good indicators of perceived-relevance for textual documents.

Research on non-textual relevance assessment have also used the approach of aggregated features.
For instance, relevance of images have been studied in \cite{zhang2010EyeMovementInteraction, hardoon2010ImageRankingImplicit, klami2008CanRelevanceImages, brouwer2009AreYouReally,golenia2015LiveDemonstratorEEG, golenia2018ImplicitRelevanceFeedback, hajimirza2010FindingUserInterest},
while that of live webpages were studied in \cite{loyola2015CombiningEyeTracking, gwizdka2015DifferencesEyeTrackingMeasures, wu2019InvestigatingRoleEye}. 
Though most studies used aggregate features for the whole stimuli duration, the authors of \cite{gwizdka2017TemporalDynamicsEyetracking} report that features from two-second windows near the end of viewing had more discriminating power than those obtained near the beginning of viewing.
Thus, collapsing eye-tracking data and thereby losing temporal information, results in our reduced understanding of human relevance assessment.

In terms of models used, most studies employed popular classifiers like Random Forests (RF) and Support Vector Machines (SVM).
Few studies employed Hidden Markov Models \cite{simola2008UsingHiddenMarkov} and  Neural Networks \cite{chow2015ClassifyingDocumentCategories}. 
Performance was varied, based on the choice of features.
For instance, Wu et al. \cite{wu2019InvestigatingRoleEye} predicted user-satisfaction while examining search results. They used advanced mathematical features (e.g., max. and SD of integrated curvature of fixations, using Frenet frame and Bishop frame) which are usually difficult to conceive in information science research. 
They obtained F1 scores in the range of 0.5 - 0.7 using RF and SVM.
Slanzi et al. \cite{slanzi2017CombiningEyeTracking} predicted web-surfer's click-intention from eye-tracking features.
They used a battery of classifiers, but the F1 scores were not promising. 
Thus, appropriate feature selection is crucial to obtain good prediction performance when aggregating eye-tracking data.

Summarily, we see that use of aggregated eye-tracking features and traditional classification techniques resulted in unpromising performances for relevance prediction.
While statistical tests were significant at the $p < .01$ level, the classification and prediction accuracies were rarely more than 70\% \cite{slanzi2017CombiningEyeTracking, wenzel2017RealtimeInferenceWord, simola2008UsingHiddenMarkov, gwizdka2015DifferencesEyeTrackingMeasures}.
In our proposed method, we demonstrate that utilizing the entire eye-tracking data, and applying image classification technique, we can predict perceived-relevance with up to 80\% accuracy.


\vspace{-1em}
\subsection{Eye-movement Scanpath Analysis}
\label{sec:bg_et_scanpath}

The issues discussed in Section \ref{sec:bg_relevance_et} arise from the dearth of appropriate analysis methods for eye-tracking data.
The entire eye-movement trajectory of a user on a stimulus is called a \textbf{scanpath}.
A scanpath has various spatial and temporal attributes associated with it: its geometric shape and size, count and duration of fixations, and the sequential information of the fixations as they occurred in time.
As of this writing, we do not have a standard loss-less method for representing all this information into a set of features.
Analyzing the differences between groups of scanpaths on relevant and irrelevant documents becomes tricky, and the results vary based on the chosen set of features.

Several scanpath comparison algorithms have been proposed, which either 
use 
(a) the actual fixation points from the eye-movement trajectory \cite{jarodzkaVectorbasedMultidimensionalScanpath2010, dewhurstItDependsHow2012}, or,
(b) a string representation of the trajectory, using letter-labels to categorize each fixation \cite{andersonComparisonScanpathComparison2015, holmqvist2011eye}.
The first approach works only with scanpaths having equal number of fixations.
To deal with scanpaths having diferring number of fixations, the algorithm deletes or clusters some fixations together (simplification step) such that all scanpaths have identical number of fixations.
We argue that such an approach may work well for non-reading tasks (e.g. viewing images), but for analyzing eye-movement while reading, all fixation points should be preserved.
Nearby fixations on different distinct words should not be clustered together into one fixation, as they may contain important information pertinent to the reading task.
The second scanpath comparison approach uses a string representation of the two scanpaths, and compares them using either the Levenshtein distance \cite{brandt1997spontaneous, duchowski2010scanpath} or the Needleman-Wunsch algorithm \cite{cristino2010scanmatch, west2006eyepatterns}.
This method assumes that annotated data is available for all the fixations. 
However, such annotations are not available when we do not have pre-existing insights about the eye-movements for our task.
A common limitation of both the methods is that they work for pairwise comparisons only, and cannot be easily extended to compare between groups of scanpaths.

\subsection{Image Classification using Convolutional Neural Networks (CNN)}
\label{sec:bg_cnn}
As introduced in Section \ref{sec:intro}, we propose an image-classification approach for predicting perceived-relevance. 
Over the last decade, image classification, and computer vision in general, has seen tremendous improvement by starting to re-use the Convolutional Neural Network (CNN).
Although developed in the 1970s, CNNs did not play a major role in computer vision research until 2010s, due to lack of adequate computing capabilities for fast execution.
In 2012, Ciresan et al. \cite{cirecsan2012multi} applied max-pooling operation after convolution, using dedicated hardware GPUs.
This process significantly improved the benchmark performances of numerous computer vision algorithms.
Around the same time, the \textbf{ImageNet} Large Scale Visual Recognition Challenge (ILSVRC) \cite{russakovsky2015imagenet} began to be organized annually.
The goal of the challenge was to beat previous years' top-performances for object recognition tasks\footnote{object recognition encompasses image classification and object detection} on more than 14 million annotated images.
Various research institutions began participating in the challenge, and the competition spearheaded the emergence of high-performing CNN architectures that began to be regarded as benchmarks.
Examples of such benchmark architectures are VGG \cite{simonyan2014very}, DenseNet \cite{huang2017densely}, ResNet \cite{he2016deep, he2016identity}, Inception \cite{szegedy2015going, szegedy2016rethinking} and InceptionResNet (combination of Inception and ResNet architectures) \cite{szegedy2017inception}\footnote{The architecture names often have numeric suffixes to denote the number of hidden layers. E.g., VGG16, VGG19 DenseNet121, DenseNet201, etc.}.

An interesting feature of CNN based image-classifiers is that the `knowledge' learnt by the network for solving one problem can be reused to solve another related problem. 
This is called \textbf{transfer learning}.
The initial layers of a CNN based image classifier learns low-level image-features (edges, shapes, and corners), while the final layers learn increasingly abstract and task specific features \cite{yosinski2015understanding, zeiler2014visualizing, krizhevsky2012imagenet}.
Since low-level image-feature detection is required in all forms of automated image-understanding, transfer learning works well for research problems having relatively low-sized datasets.
For this reason, popular deep-learning frameworks (e.g., Keras, PyTorch etc.) include many benchmark CNN architectures, with their weights pre-trained on the ImageNet challenge.
In this work, we utilize several such benchmark CNN image classifiers to predict the perceived-relevance of documents from scanpath images.
%

A CNN is often considered as a ``black-box'', because its inner working are not easily understandable.
Various methods have been proposed to understand why the network makes a particular prediction \cite{springenberg2014striving, selvaraju2017grad}.
One such method is Gradient-weighted Class Activation Mapping (Grad-CAM) \cite{selvaraju2017grad}.
The Grad-CAM method produces a heatmap, which is similar to an attention map, and highlights the regions of the input image that was focused on for making the prediction.
For `known` research problems (e.g. detecting cats vs. dogs in images) this visualization helps to understand whether the CNN is paying attention to the relevant image regions.
In our case of classifying scanpath images according to perceived-relevance, the Grad-CAM visualizations can offer new insights about human reading behaviour on relevant and irrelevant documents.


Examining the challenges involved in relevance prediction using eye-tracking data (Section \ref{sec:bg_relevance_et} and Section \ref{sec:bg_et_scanpath}), and also the opportunities provided by image classification and transfer learning (Section \ref{sec:bg_cnn}), we propose an image-classification based solution for the problem of perceived-relevance prediction.
The advantages of our method are:

\begin{enumerate}
    \item unlike previous studies where eye-tracking data was collapsed into a set of single-number features, our method allows to use all the data points that are collected, for a more nuanced analysis
    
    \item the spatial and temporal characteristics of eye-movement scanpaths can be utilized to make inferences 
    
    \item our method is content independent, and does not require knowledge of what the user is viewing on the screen
    
    \item unlike approaches in reading-related studies, our method does not require additional insights about the data (e.g. need not label fixations as reading, scanning, etc.)
\end{enumerate}


\section{User Study}
\label{sec:user_study}

\subsection{Experimental Design and Procedure}
\label{sec:exp_design_proc}

A controlled lab experiment was conducted in the Department of Kinesiology, University of Maryland, College Park. Participants ($N=25$, college-age students) judged the relevance of short news articles for answering a trigger question. Eye-tracking and EEG signals were recorded. 
In this paper, we report a novel analysis using only the eye-tracking data. 

\begin{figure}[!htbp]
\centering
\includegraphics[clip=true, trim=0 450 550 0, width=\linewidth]{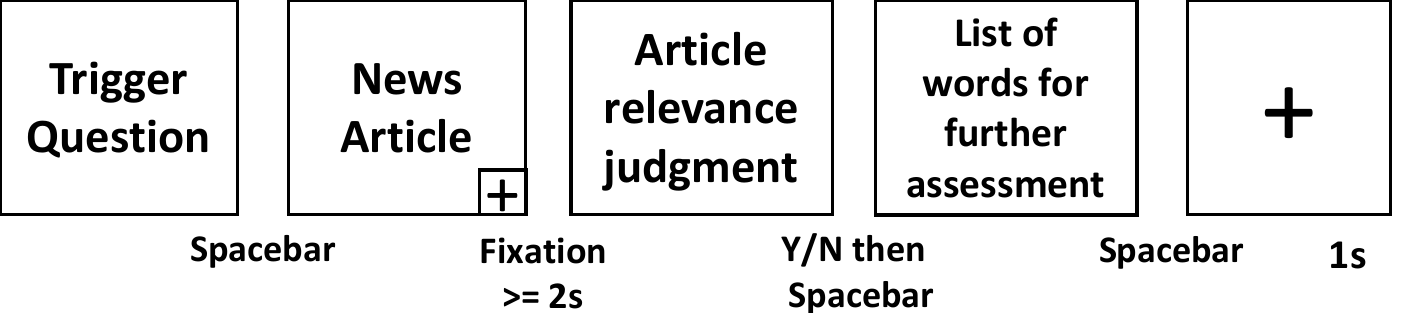} 
\caption{
One trial in the experimental procedure.
}
\label{fig:exp_proc}
\end{figure}

The main element of the experimental procedure was a trial (Figure \ref{fig:exp_proc}). 
In each trial, a trigger question was shown first. 
The trigger questions was a short, one-sentence question, informing participants what to look for in the subsequently presented documents (e.g. ``What is the birth name of Jesse Ventura?''). 
After the trigger question, a short news article was displayed, then a text relevance response (Y/N) screen appeared, then a list of words for further assessment was shown. 
Participants progressed between stimuli by pressing a space bar, with an exception of moving from a news article to the text-relevance response screen, which occurred by participants fixating their eyes for two seconds or longer in the lower-right screen-corner to indicate their readiness for relevance judgement. 
Finally, a fixation screen was shown for one second between trials.
The list of words for further assessment are not analysed in this paper.
The news articles were chosen to have three levels of relevance with respect to the trigger question: 
\begin{itemize}
    \item \textbf{Relevant (R)}: the article explicitly contained the exact answer asked in the question
    
    \item \textbf{Topical (T)}: partially relevant --  the article did not contain the exact answer to the question, but was on the topic of the information asked in the question
    
    \item \textbf{Irrelevant (I)}: did not contain the answer to the question
\end{itemize}
\noindent
We regard this three-level relevance for each news article as the article's \textbf{document-relevance}. 
The source of these relevance labels are discussed in Section \ref{sec:stimuli}.

There were 40 trigger questions; each of them was associated with three news articles designed to contain exactly one R document, and one or more T or I documents.
Thus, the following 12 permutations were possible for each question: 
\texttt{\{RTI, RIT, RTT, RII, TRI, IRT, TRT, IRI, TIR, ITR, TTR, IIR\}}. 
This yielded 120 sets of question + news article; each set constituting one experiment trial. 
The order of trials was randomized for each participant to mitigate order effects. 
Participants rated the news article as relevant or irrelevant (by pressing Y or N key) based on their judgement of whether the article contained an answer to the trigger question.
These binary responses from each participant, for each news article, are regarded as the \textbf{perceived-relevance} for the user-document pair.
Participants performed a training task consisting of six trials (two questions and six documents) before the 120 trials.

\vspace{-1em}
\subsection{Stimuli Dataset}
\label{sec:stimuli}
The set of 40 trigger questions were selected from the TREC 2005 Question Answering Task \cite{voorhees2003overview}.
The collection of 120 short news articles and their document-relevance labels came from the AQUAINT Corpus of English News Text \cite{graffaquaint} (the same collection used in TREC 2005 Q\&A Task).
The news articles were carefully selected to have nearly similar text-length (mean length: 178 words, SD: 30 words).

\subsection{Apparatus}
\label{sec:apparatus}
Eye-tracking data was recorded in the lab on a Windows laptop PC connected to an SMI RED250 eye-tracker. Participant relevance responses were recorded on a remote server. 
The eye-tracker sampling rate is 250 Hz, and an accuracy up to $0.4^o$ of visual angle.
The screen resolution was $1680 \times 1050$. Eye-tracking data was captured  by SMI iViewX software and the stimuli were presented by SMI Experiment Center 360 v3.0 software. 
The textual stimuli were entered to Experiment Center’s text editor as the text elements, and displayed in black Times font on a light-grey background.
Line-height was approximately 32 pixels.

\begin{figure}[!htbp]
\centering
\includegraphics[clip=true, trim=5 10 5 5, width=0.9\linewidth]{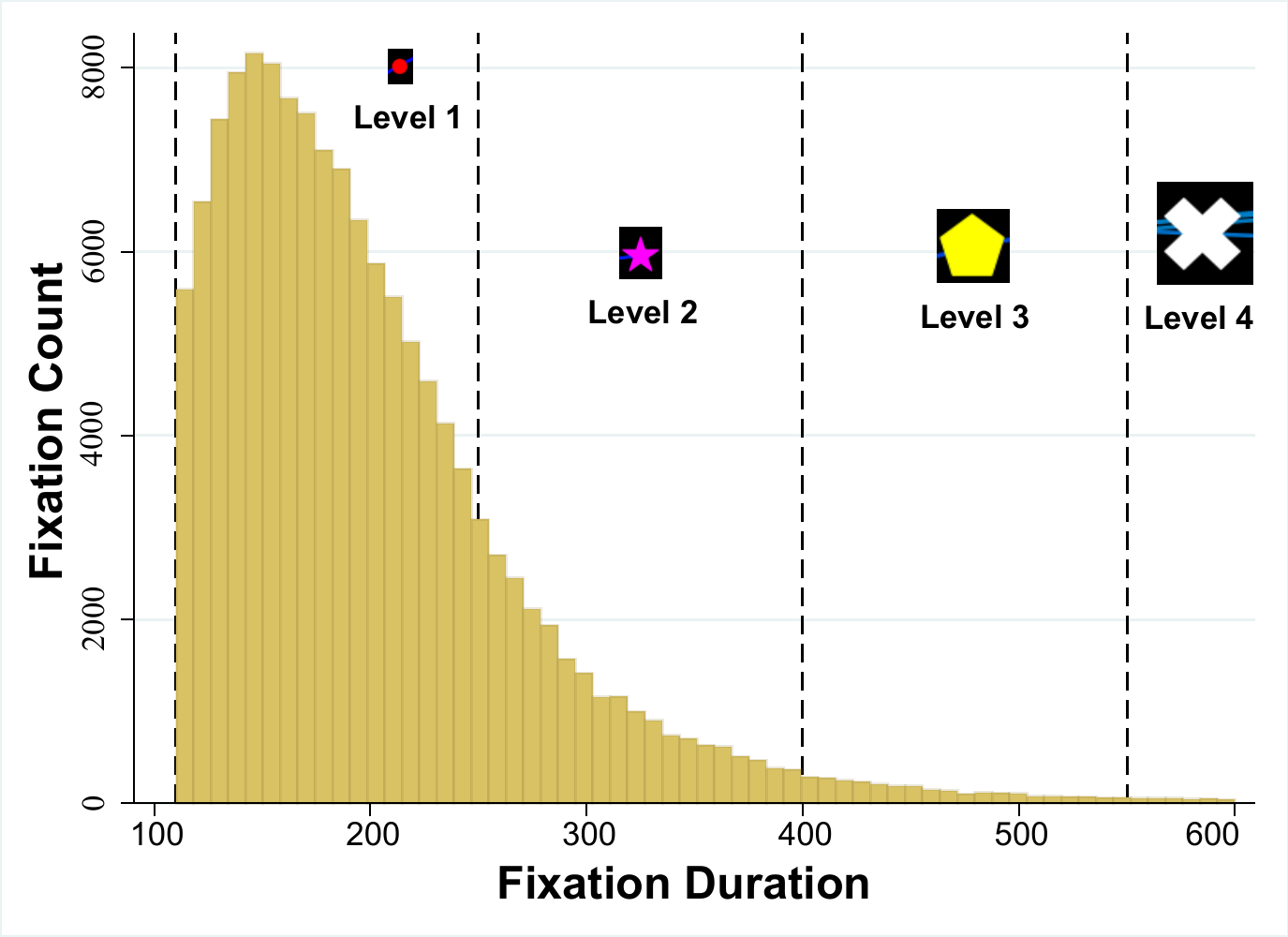} 
\caption{
Distribution of recorded fixation-durations, and the corresponding encoding marker for representing fixations belonging to different levels (See Section~\ref{sec:encoding_scanpaths_fixations}).
}
\label{fig:fixn_dur_hist}
\end{figure}

\section{Data Analysis}
\label{sec:analysis}

Eye-tracking data was processed using the SMI BeGaze: Analysis Software (version 3.2). 
Data recording for one participant failed, hence we report analysis for ($N=24$). 
Fixations were detected using Velocity-Threshold Identification (I-VT) algorithm, as implemented in the BeGaze software, with default parameter values.


\begin{figure*}[!thbp]
\centering
\includegraphics[clip=true, trim=0 155 60 0, width=\linewidth]{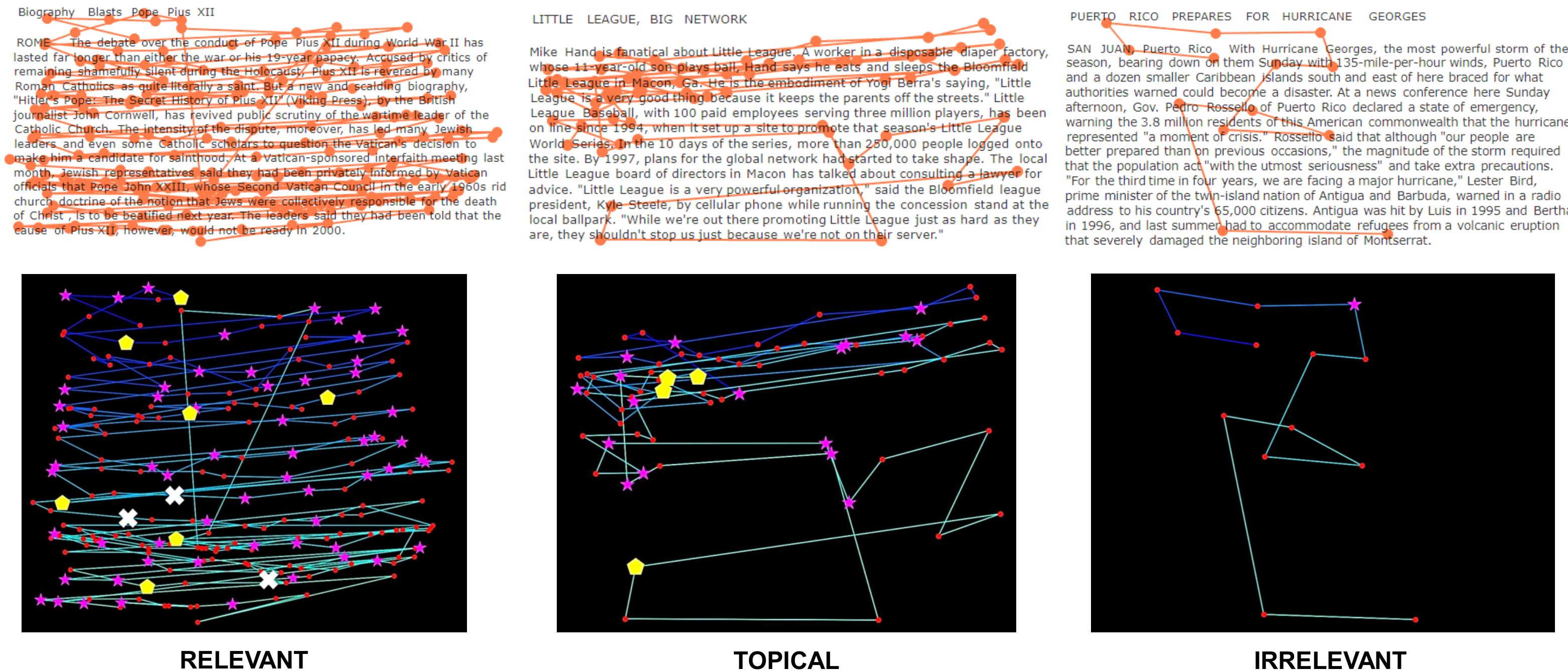} 
\hfill
\caption{
Top: Typical eye-movement patterns when reading relevant, irrelevant, and topical documents.
Bottom: Examples of generated scanpath images, which are used to train CNN classifiers for predicting the user's perceived-relevance of the documents.
This figure is best viewed in colour, on screen.
}
\label{fig:scanpath_encoding}
\end{figure*}

\vspace{-1em}
\subsection{Generating Scanpath Images}
\label{sec:encoding_scanpaths}
We generated scanpath images from eye-tracking data of user-document pairs, using only three attributes of eye-fixations:
screen-coordinates (in pixels), fixation duration (in ms), and start time of the fixation relative to stimulus-onset.
We used Python Matplotlib library \cite{hunter2007matplotlib} to generate the scanpath images.
CNNs have been shown to be good at detecting local patterns within images \cite{ANDREARCZYK201795, SRINIVAS201725}.
Since we were preparing the images for training a CNN classifier, we made the following design choices: 

\subsubsection{\textbf{Fixations:}}
\label{sec:encoding_scanpaths_fixations}
Eye-fixations were encoded as marker points having varying shapes, sizes, and colours. 
These were controlled by the fixation duration as follows:
\begin{itemize}
    \item 110 - 250 ms: Level 1 fixations, encoded as red circle
    \item 250 - 400 ms: Level 2 fixations, encoded as pink star
    \item 400 - 550 ms: Level 3 fixations, encoded as yellow pentagon
    \item > 550 ms: Level 4 fixations, encoded as white cross
\end{itemize}
These levels were identified empirically. 
We examined the distribution of fixation durations in our data, and roughly divided the range into three equal partitions (Figure \ref{fig:fixn_dur_hist}).
Fixations having durations less than 110 ms were discarded \cite{widdel1984operational, Salvucci_2000_IFS_355017_355028}.
The marker-size was made to increase with the Level number.
The fixation markers were chosen to be grossly different from each other (instead of, say, only circles), so that the CNN could possibly identify spatial patterns of similar-duration fixations.

\subsubsection{\textbf{Linearized Saccades:}}
Saccades are rapid eye-movements between two fixation points.
Ideally, they follow ballistic paths.
To keep things simple for our analysis, we plotted linearized saccades: the effective eye-movement between two fixations, represented as a straight line connecting the two points.
For brevity, henceforth we will say `saccade' to mean `linearized saccade'.
We controlled the colour of the saccade lines to follow a linear colour scale, based on their temporal occurrence (`Winter' colourmap in Matplotlib\footnote{\url{ https://matplotlib.org/3.1.1/gallery/color/colormap_reference.html}}).
The colour of the saccades changed linearly from blue (first saccade) to green (final saccade).
Each individual saccade had a solid colour.

We also tested controlling the width of the saccade lines using saccade velocity (ratio of screen-distance covered to time taken). 
However, doing so made the scanpath-image too crowded, especially for scanpaths having more than 50 fixations.
So we kept the width of the saccade lines constant at 2 pixels.

\subsubsection{\textbf{Colours:}}
Care was taken to select the colours of the fixations and the saccades. 
Using a colour wheel, the colours of the different fixation markers were chosen to be far apart, from each other, as well as from the range of colours used to draw the saccades.
We hypothesized that these colour choices would enable the CNN classifier to easily distinguish between fixations and saccades, and identify necessary patterns. Examples of typical eye-movement patterns on three types of documents, and their corresponding generated scanpath images are shown in Figure \ref{fig:scanpath_encoding}.

\subsection{Machine Learning Setup}
\label{sec:ml_setup}

Data was available for 24 participants, where each participant judged the binary relevance of 120 news articles.
In total we had eye-tracking data for 2,880 user-document pairs, or 2,880 scanpaths.
After data cleaning, we decided to use scanpaths having 10 or more fixations.
We assumed that at least 10 fixations, or a minimum eye dwell-time of 1 second on the document (at 100 ms / fixation) is required to make a relevance assessment.
This left us with 2,579 scanpath images.

\subsubsection{\textbf{Train / Validation / Test Partition:}}
As \textit{human}-information-interaction researchers, we are more interested in studying human behaviour.
So we used the participants' perceived-relevance labels as the ground-truth for our classification task (and not the document-relevance obtained from TREC dataset).
Out of the 2,579 scanpath images, only 806 (31.2\%) were for documents marked relevant.
Thus, there was almost a 1:2 class imbalance.
Since this is an initial attempt to apply image classification on scanpath images, we decided to use a balanced dataset. 
So we randomly sampled 806 images from the pool of irrelevant scanpath images, and created a perfectly balanced dataset of 1,612 images.
We used an approximate 60-20-20 split to randomly place \textbf{966 images in the training set, 314 images in the validation set, and 332 images in the test set}.
The relevant/irrelevant class balance was preserved in each set.
All random selections were performed using the MySQL \texttt{rand()} function.

\vspace{-1em}
\subsection{Analysis Procedure}
\label{sec:analysis_proc}

\subsubsection{\textbf{Image Classification Setup:}}
\label{sec:analysis_cnn}
We posed our binary classification problem as follows: 
\textit{given \textbf{only} the scanpath image of a user's eye movements on a short news article, did the user perceive the article to be relevant for answering a trigger question?}

For this binary classification problem, we analysed the performance of six popular CNN based architectures:
VGG16 and VGG19 \cite{simonyan2014very}, 
DenseNet121 and DenseNet201 \cite{huang2017densely}, 
ResNet50 \cite{he2016deep, he2016identity}, 
and
InceptionResNet (version 2) \cite{szegedy2017inception}.
All the architectures had benchmark performances in the ImageNet challenge \cite{krizhevsky2012imagenet}. 
To examine whether the obtained results were reproducible in different environments and software versions \cite{crane2018questionable}, we ran the analyses independently on two popular Python deep-learning frameworks: TensorFlow-Keras\footnote{\url{https://www.tensorflow.org/guide/keras}}, and PyTorch-fastai. 
The architecture of the TensorFlow-Keras implementation was: 

\vspace{0.5em}
{\small
\noindent
\textbf{CNN model} (initialized with pre-trained ImageNet weights) --> \textbf{Fully Connected Layer} (256 nodes, ReLU activation, with/without L1L2 regularization) --> \textbf{Dropout} (probability=0.2) --> \textbf{Fully Connected Layer} (1 node, Sigmoid activation).
\hspace{0.5em}
Optimizer: Stochastic Gradient Descent (SGD)
}
\vspace{0.5em}

\noindent
In PyTorch-fastai, we built the classifier using the \texttt{cnn\_learner} module\footnote{\url{https://docs.fast.ai/vision.learner.html\#cnn_learner}}, which initializes the model with random weights, and trains from scratch.
We ran the TensorFlow-Keras implementation on FloydHub GPU Cloud Server\footnote{\url{https://www.floydhub.com}} 
(NVIDIA Tesla K80 GPU, 12 GB memory, 61 GB RAM),
and the PyTorch-fastai implementation on Google Colab\footnote{\url{https://colab.research.google.com}} 
(NVIDIA Tesla T4 GPU, 15 GB memory, 26 GB RAM).

We trained the models on the training set, and used the validation set for very basic hyper-parameter tuning (learning rate, number of epochs, optimizer momentum, etc.).
Since our intention was to see whether the method works, and not to obtain the best benchmark performance, we performed minimal hyper-parameter tuning. 
Finally, we took the best set of models obtained after tuning the hyper-parameters \textbf{(epochs: 6, batch-size: 16, momentum: 0.9)}, and used them to predict the labels of the test set.
The top portion of Table~\ref{tab:results_table} reports the results from the TensorFlow-Keras implementation, while Table~\ref{tab:results_fastai} reports the results from the PyTorch-fastai implementation.
The discussions are centred around the results from the TensorFlow-Keras implementation.

\vspace{-0.75em}
\subsubsection{\textbf{Comparison to Existing Standard:}}
\label{sec:analysis_trad_ml}
We also compared our method to existing approaches for inferring relevance using eye-movements, where the data is collapsed into a set of handcrafted features (discussed in Section~\ref{sec:bg_relevance_et}). Perceived-relevance of documents are predicted from these features using popular classifiers like Random Forests \cite{wu2019InvestigatingRoleEye, jimenez-molina2018UsingPsychophysiologicalSensors} and Support Vector Machines (SVM) \cite{slanzi2017CombiningEyeTracking, li2018UnderstandingReadingAttention}.
We computed 20 such hand-engineered features, aggregated at the user-doc level, and classified them using Random Forest and SVM.
This analysis was done using Python Scikit-learn library.
Similar to our approach with the CNN classifiers, we started with the default hyperparameter values of the Random Forest and the SVM classifier from the Scikit-learn library, and then performed basic parameter tuning. Finally, we selected the best performing models.
The bottom portion of Table~\ref{tab:results_table} reports these results.
The handcrafted features are discussed in Section~\ref{sec:res_comp_ml}.


\begin{table*}[!thbp]
\caption{
Performances of two different methods to predict perceived-relevance from eye-movements, ordered by decreasing F1 score for the Test Set. 
\textit{Top:} CNN classifiers on scanpath images.
\textit{Bottom:} traditional classifiers on aggregate features. 
}
\label{tab:results_table}
\centering
\includegraphics[clip=true, trim=0 110 601 0, width=0.65\linewidth]{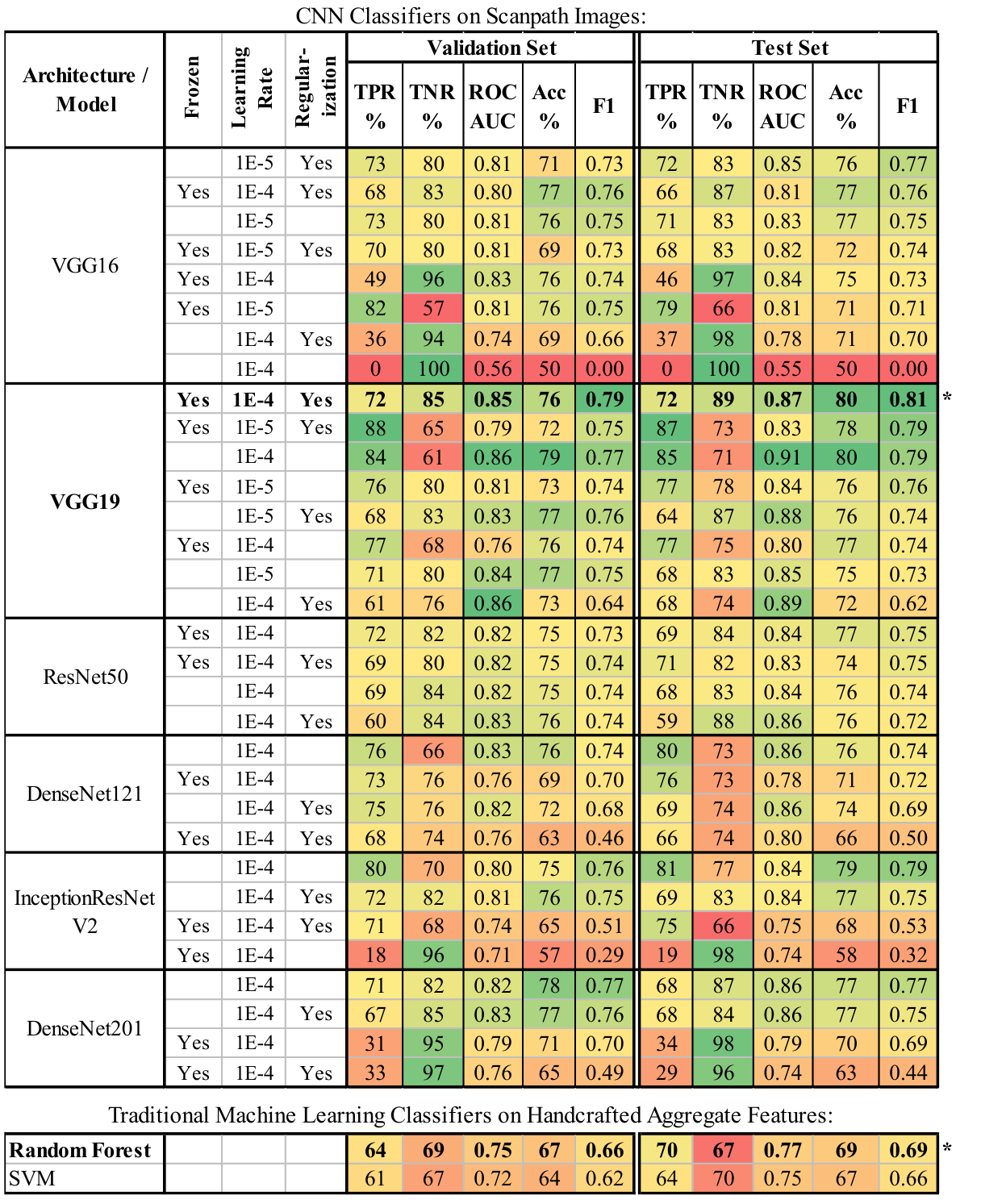} 
\hfill
~\\
\footnotesize{
\begin{flushleft}
Colour scales rank the performance of each row from best (green) to worst (red), across both methods.
Asterisk (*) indicates best performance for each method.
~\\
For CNN classifiers:
\textbf{Frozen:} if Yes, then weights of the CNN layers (pre-trained on ImageNet) were frozen during training.
\textbf{Regularization:} if Yes, then L1 and L2 regularization with decay = 0.01 was used in the Fully Connected Layer
(See Section \ref{sec:analysis_cnn} for neural network architecture).
\end{flushleft}
}
\end{table*}

\section{Results \& Discussion}
\label{sec:results_disc}


\begin{figure*}[!thbp]
\centering
\includegraphics[clip=true, trim=0 95 190 0, width=0.9\linewidth]{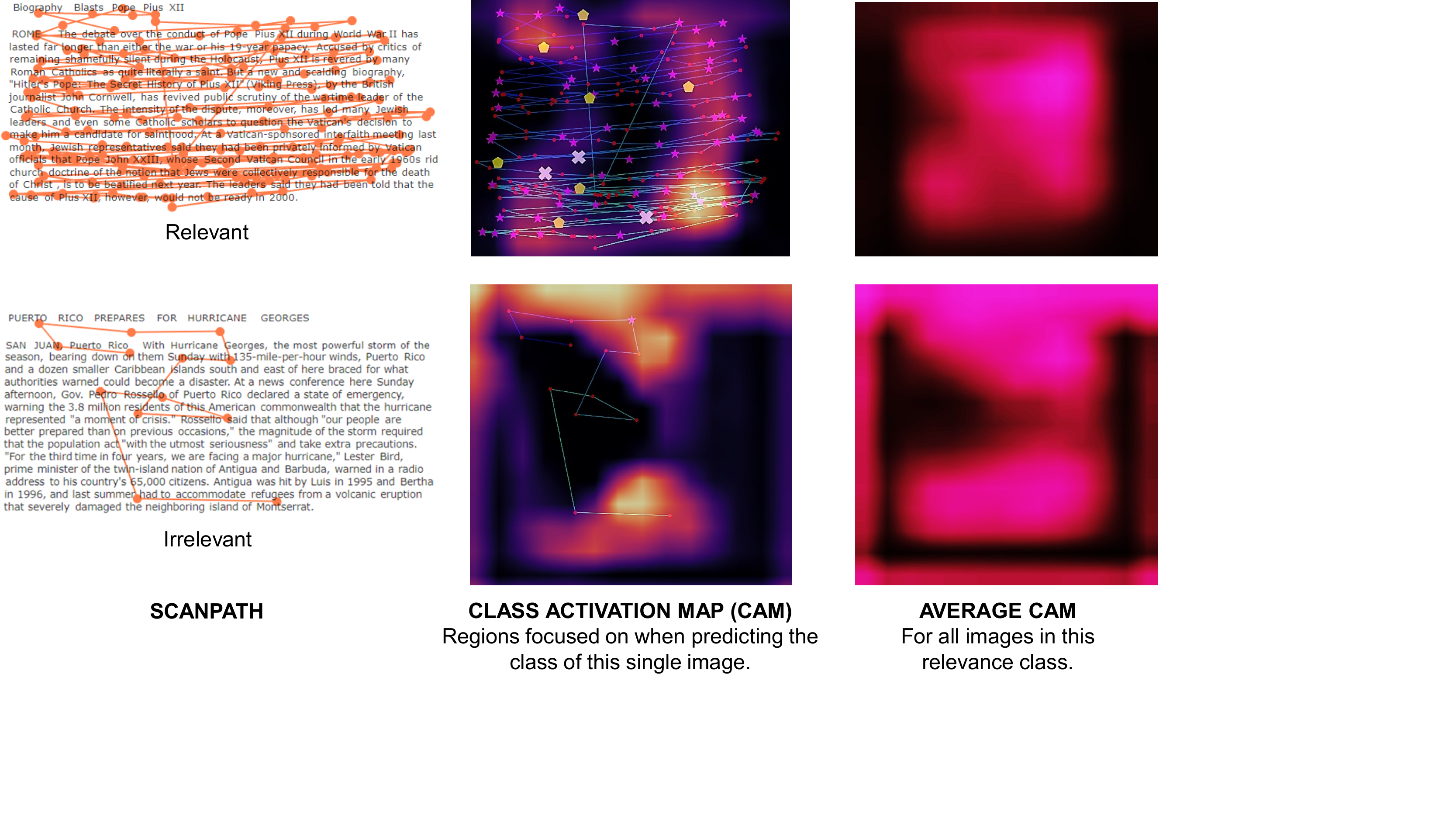} 
\hfill
\caption{
Attempt to interpret how the CNN classifiers made predictions.
Middle column shows heatmaps obtained using Grad-CAM technique for a single image.
Right column shows average of all such heatmaps.
All heatmaps are generated using the best performing model and hyperparameters from Table~\ref{tab:results_table} (VGG19, F1: 0.81).
Inferences are discussed in Section \ref{sec:res_interpretability}.
}
\label{fig:class_activation_map}
\end{figure*}

\subsection{Scanpath Image Classification}
\label{sec:res_classifier}

We report the performance of our proposed scanpath image classification method, by testing six different CNN classifier architectures (Table~\ref{tab:results_table}, top).
To easily compare our results to those reported in previous papers, 
we report five different metrics: 
percentages of correct predictions for both relevant and irrelevant documents, as True Positive Rate (TPR \%) and True Negative Rate (TNR \%);
accuracy (Acc \%); 
area under the ROC curve (ROC AUC); and F1-score (F1).
TPR and TNR are also known as sensitivity and specificity, respectively.
We have ranked the image classifiers according to their F1 scores on the Test Set.
We have taken care to report all range of performances -- best, average, and worst -- to provide realistic expectations of using this method.

From Table~\ref{tab:results_table}, we have the following observations: 
First, all the classifiers have comparable F1 scores on both the Validation Set and the Test Set.
Despite having less than 1000 training images (which is quite low by deep-learning standards), the models did not overfit, and generalized well on the unseen Test Set.
Second, VGG16 and VGG19 architectures show the best performances (seven out of the top-10 F1 scores).
These are ``shallow'' networks, having 16 and 19 layers respectively.
Very deep models (e.g. DensNet201 or ResNet50, with 201 and 50 layers, respectively) on the other hand, occur once each within the top-10 F1 scores.
Thus, shallower models performed better for relevance-prediction from scanpath images, than deeper models.
Third, frozen versions of shallower models performed better than their unfrozen counterparts, while unfrozen versions of deeper models had better F1 scores than their frozen versions.
When the models were trained in frozen mode, only the fully connected layers had their weights updated by gradient descent, while the weights of the CNN layers were kept frozen (refer to Section \ref{sec:analysis_cnn} for model architecture).
Shallower models therefore effectively re-utilized the training received from another object classification task (i.e. the ImageNet challenge \cite{krizhevsky2012imagenet}), while the deeper models needed to learn new weights to have similar performance.
Fourth, from Table~\ref{tab:results_fastai} we see that F1 scores of the same architectures implemented in PyTorch-fastai are similar to those obtained using the TensorFlow-Keras implementation.
The results are thus reproducible across different libraries and software environments.

Accordingly, using latest CNN classifiers, comparatively less training data, and leveraging the power of transfer-learning, it is possible to predict the perceived-relevance of documents from scanpath images 
with F1 score up to 0.81, and up to 80\% accuracy (Table~\ref{tab:results_table}, VGG19 row, marked with asterisk).

\begin{table}[!htbp]
\caption{
Results (F1-scores) from PyTorch-fastai implementation, with similar configurations as in Table~\ref{tab:results_table}. 
}
\label{tab:results_fastai}
\includegraphics[clip=true, trim=0 425 775 0, width=0.55\linewidth]{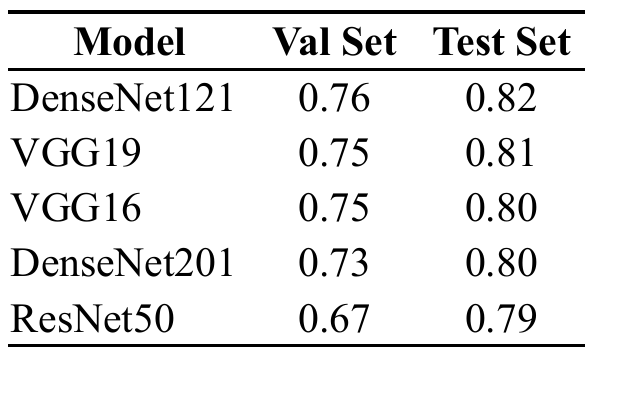} 
\vspace{-2em}
\end{table}

\vspace{-1em}
\subsection{Comparison with Traditional Classifiers}
\label{sec:res_comp_ml}


To compare our result to the existing approaches of today, 
we tested the performance of two popular classifiers --  Random Forest and SVM -- using 20 handcrafted features informed by literature (Table~\ref{tab:results_table}, bottom).
The highest Test Set accuracy obtained was 69\% (our proposed method achieves 80\%), and the highest F1-score obtained was 0.69 (our method achieves 0.81).
The five most important features, as obtained from the Random Forest classifier, were 
(1) vertical scan speed,
(2) HV ratio (ratio of total horizontal movement to total vertical movement, normalized by screen dimensions),
(3) SD of fixation durations,
(4) mean saccade length, and
(5) task duration.
To make this comparison fair against our proposed method, we had included the counts of fixations in the different levels (1-4) -- that we encoded with special markers in the scanpath images (Section \ref{sec:encoding_scanpaths_fixations}) -- in our handcrafted feature set. 
However, those level-wise fixation counts were placed among the ten least-important features by the Random Forest classifier.
Thus, the scanpath image classification method performs much better than using handcrafted features.

\vspace{-1em}
\subsection{Comparison with Related Works}
\label{sec:res_comp_related_work}
Our method vastly improves upon performance-measures reported in literature on related work. 
We first discuss the studies that had similar experimental setups as our own, and also predicted relevance from eye-tracking data.
Our best performing classifier surpasses the numbers reported in these studies by at least five percentage points on average, w.r.t. ROC AUC, Accuracy, and/or F1 score.
For instance, 
Chow et al. \cite{chow2015ClassifyingDocumentCategories} predicted document-categories from eye-movement data of analysts, using a neural network classifier.
They reported an accuracy of 70\%, but did not mention the specific eye-tracking features used.
Wenzel et al. \cite{wenzel2017RealtimeInferenceWord} inferred the relevance of individual words from fixation duration and EEG features. 
They reported that only using fixation-duration gave an AUC of 0.51 (marginally better than chance), whereas combining fixation-durations with EEG features improved the AUC to 0.63. 
However, they did not state the kind of classifiers used.
Compared to the above figures, our best performing model -- VGG19, by Test Set F1 score -- has an ROC AUC of 0.87, accuracy of 80\%, and F1 score of 0.81.
Gwizdka \cite{gwizdka2014CharacterizingRelevanceEyetracking} predicted relevance of short documents, and he reported a maximum accuracy of 74\% using decision trees.
However, some of the features he used were content-dependent (number of fixations on words, count and percentage of words fixated upon, etc.). Our method, on the other hand, is both content and task independent.

In our literature search, the only study found to have comparable and better performance than our image classification method, for a similar relevance judgement task, was reported in \cite{gwizdka2017TemporalDynamicsEyetracking}.
Employing proximal SVM as the classifier, the best performance using only eye-tracking features were reported to have an AUC of 0.95, and accuracy of 86\%.
However, their approach had two distinct differences from our method:
First, all fixations were passed through a two-stage reading model to label them as reading or scanning.
Then separate features were calculated for the groups of reading and scanning fixations.
In contrast, our method simply takes all the raw fixations and encodes them directly into the scanpath image.
There is no need for pre-classification, which requires additional insights about the data.
Second, the classification features were computed using windows of 1 second and 2 seconds near the beginning, middle, and end of the reading trials.
Higher prediction accuracies were obtained using the values of end-window, than values of the beginning-window.
Differently, our method considers the entire duration of the reading trial for the prediction task.

We now discuss classification and prediction results from other related yet different studies, which employed eye-tracking in the domain of interactive IR.
Simola et al. \cite{simola2008UsingHiddenMarkov} predicted task-category (word search, question answer, or reading by interest) from the scanpath, and they obtained 59.8\% accuracy using logistic regression on fixation count, mean and SD of fixation durations, and mean and SD of saccade length.
Slanzi et al. \cite{slanzi2017CombiningEyeTracking} predicted the click-intention of web users.
Though they initially considered using eye-tracking features, those were later discarded using Random Lasso feature selection, and  EEG and pupillometry features were mainly used.
They employed a variety of classifiers, including SVM, neural network, and Logistic Regression.
However, the highest F1 score obtained was 0.4, using the neural network classifier. 
Although they reported the highest accuracy of 71\% for Logistic Regression, it had low precision and recall, and thus low F1 score of 0.33.
Gwizdka et al. \cite{gwizdka2015DifferencesEyeTrackingMeasures} predicted visits and revisits to relevant and irrelevant webpages, using fixation-duration, saccade-duration, and saccade-length.
They reported a maximum accuracy of 61\% using Flexible Discriminant Analysis.
Though our prediction problem was somewhat different than the ones discussed above, we hypothesize that our method can possibly obtain better performances on these prediction problems as well, since the scanpath image classifier did not receive any information about the task.





\vspace{-1em}
\subsection{Interpreting Reasons for Prediction}
\label{sec:res_interpretability}

In this section, we attempt to interpret how the best performing CNN classifier (VGG19, Test set F1 score: 0.81, from Table~\ref{tab:results_table}) made predictions.
We employed Gradient-weighted Class Activation Mapping (Grad-CAM) \cite{selvaraju2017grad} for this purpose.
Given a scanpath image, the Grad-CAM technique produces a heatmap (Class Activation Map) indicating which pixels in the image are considered important for making a prediction.
This is similar to feature-importances in Random Forests, but is specific to each scanpath image.
Examples of such heatmaps are shown in the second column of Figure \ref{fig:class_activation_map}.
To understand whether the CNN had identified some patterns about human reading behavour on relevant and irrelevant documents,
we generated an average heatmap for each class, using all the scanpath images in the Test Set. 
These average heatmaps are shown in the third column of Figure \ref{fig:class_activation_map}.

For the heatmap of the relevant scanpath image, the classifier appears to have focused more on the right side of the scanpath (Figure \ref{fig:class_activation_map}, second column).
This can be explained by findings about reading versus scanning behaviour \cite{gwizdka2014CharacterizingRelevanceEyetracking}. 
When people read relevant documents, their eyes move more horizontally than vertically. 
They also continue to read till the end of every line, and then move on to the next line.
Therefore, fixations occur at the end of most lines, in a consistent manner from top to bottom.
Specific to our task, the participants possibly read the news article headlines first, to quickly decide if the answer to the trigger question could be found in the article. 
When the headline looked relevant, they continued to read into the body of the article, and read till the end of every line (Figure \ref{fig:class_activation_map}, Relevant Scanpath).

In the irrelevant scanpath heatmap, the classifier possibly focused on the top and bottom regions instead (Figure \ref{fig:class_activation_map}, second column).
People usually scan or skim irrelevant information, and produce more vertical eye-movements than horizontal.
Very few fixations occur near the ends of successive lines, since people rarely read irrelevant documents continuously to the ends of lines, for many lines in sequence.
For our task, the participant possibly read the headline first, similar to relevant articles.
However, when the headline appeared irrelevant, they scanned the remainder of the article in long vertical sweeps.
They may have also looked at the last few lines of the article, to search for summaries or conclusions about the article-content (Figure \ref{fig:class_activation_map}, Irrelevant Scanpath).

These patterns are further reinforced in the average heatmaps (Figure \ref{fig:class_activation_map}, second column).
In the relevant case (top), the classifier's attention was spread over a rectangular region, corresponding to the overall shape of the stimuli news-paragraphs.
The right side of the rectangle is brighter than the left, indicating that the classifier looked for fixations near line endings.
For the irrelevant case (bottom), attention is focused on islands in the top and bottom regions, with a less-focused central region.

Summarily, we hypothesize that participants initially looked at the headlines for both kinds of news articles.
For relevant articles, the headlines convinced them to read in more detail.
So they produced more fixations in the body of the article, and read till the end of every line.
For irrelevant articles, on reading the headlines (and possibly the first few lines), participants understood that the article would not be useful for answering the trigger question.
So they quickly skimmed / scanned to the bottom, and looked for concluding remarks which could solidify their initial judgement of the article.
As a result, both relevant and irrelevant scanpath images contained fixations in top region (headlines).
However, the proportion of headline-region-fixations to body-region-fixations was higher for irrelevant documents.
A similar phenomenon was reported in \cite{li2018UnderstandingReadingAttention}.
Therefore, we think the CNN classifier possibly ``learnt'' that: 


\textit{If a large number of fixations are present in the right side of the scanpath image (where line endings are located) it is probably a relevant scanpath; whereas if fixations are concentrated in the top and bottom regions, and sparse in the middle, it is possibly an irrelevant scanpath.}

\vspace{-1em}
\section{Conclusion}
\label{sec:conclusion}

In this paper, we pose the research problem of `predicting perceived-relevance from eye-movements' as a problem of `scanpath image classification'.
We employ pre-trained Convolutional Neural Networks to predict whether scanpath images correspond to reading relevant or irrelevant news articles.
The advantages of our method are:
(i) we use all of the eye-tracking datapoints available per user, and do not collapse them into features;
(ii) the spatial and temporal aspects of eye-movement scanpaths are preserved;
(iii) our method is content independent, and does not require knowledge of the content being viewed (e.g., the actual text of the news articles); and
(iv) our method does not need additional insights about the data (e.g. for labelling fixations as reading or scanning).

Our approach has several limitations.
First, since we used pre-trained image classifiers, the high resolution scanpath images ($1680 \times 1050$) were reduced to the dimensions on which the classifiers were trained on ($224 \times 224$).
This led to some loss of information, and possibly decreased the classifier's performance.
However, it is standard practice in computer vision research to downsize images, since using high- or full resolution images leads to exponentially slower execution times, and significantly more memory requirements.
Second, due to resource-limitations, we were unable to appreciably search the hyper-parameter space.
It is possible that a simpler or shallower model can achieve better performance for this task.
Third, we employed a fairly simple information search task, and used only short texts of similar type.
It is possible that more complex information search tasks on the open web can bring additional challenges.
Fourth, our participant pool was relatively homogeneous: all of them were college-age students attending the same university.

To our knowledge, this is the first attempt to approach both relevance prediction and eye-movement analysis using image classification, or more broadly, computer vision.
Our work was aimed at proof-of-concept.
We demonstrated that even with little data, this method shows promising results. 
For similar eye-tracking studies from the literature, our scanpath image classification method outperformed previously reported metrics by appreciable margins.
By looking at aggregated class activation heatmaps, we gained additional insights on how users examine relevant and irrelevant documents.
Thus, there is promising scope for improving interactive IR research, by employing computer vision algorithms in non-vision tasks.

\vspace{-1em}
\begin{acks}
The research project was funded in part by \grantsponsor{LMC}{Lockheed Martin Corporation}{}.
We thank the team from Department of Kinesiology, University of Maryland, College Park led by Professor Bradley Hatfield, and including Dr. Rodolphe Gentili, Dr. Joe Dien, and graduate students: Hyuk Oh, Kyle James Jaquess, and Li-Chuan Lo, for contributing to the experimental design, implementing it in SMI Experiment Center software, and collecting the data.
We thank Splunk Inc. for the blog post  
on using mouse trajectories for fraud detection \cite{url_splunk_blog}, which gave us the idea to adapt this approach for relevance prediction from eye-movements.
We also gratefully acknowledge the ACM SIGIR Student Travel Grant awarded to the first author.
\end{acks}

\bibliographystyle{ACM-Reference-Format}
\balance
\bibliography{chiir2020_ref1,chiir2020_ref2}
\end{document}